# LLM Bazaar: A Service Design for Supporting Collaborative Learning with an LLM-Powered Multi-Party Collaboration Infrastructure

Zhen Wu, Jiaxin Shi, R. Charles Murray, Carolyn Rosé
zhenwu@andrew.cmu.edu, jiaxins1@andrew.cmu.edu, rcmurray@andrew.cmu.edu, cprose@cs.cmu.edu
Carnegie Mellon University, PA, United States

Micah San Andres
msanandres@ucsd.edu
University of California, San Diego, CA, United States

**Abstract:** For nearly two decades, conversational agents have played a critical role in structuring interactions in collaborative learning, shaping group dynamics, and supporting student engagement. The recent integration of large language models (LLMs) into these agents offers new possibilities for fostering critical thinking and collaborative problem solving. In this work, we begin with an open source collaboration support architecture called Bazaar and integrate an LLM-agent shell that enables introduction of LLM-empowered, real time, context sensitive collaborative support for group learning. This design and infrastructure paves the way for exploring how tailored LLM-empowered environments can reshape collaborative learning outcomes and interaction patterns.

## Introduction

Providing technological support for collaborative and discussion-based learning has long been a focus in CSCL research (Gweon et al., 2006; Kollar et al., 2006; Kumar et al., 2007; Rosé and Ferschke, 2016, Naik et al., 2024). Open-source architectures like Bazaar (Adamson et al., 2014) have enabled implementation of a plethora of dynamic support interventions, even for face-to-face collaboration through multi-modal sensing (Wang et al., 2020), which can be used in a portable fashion for nearly anytime-anywhere collaboration support (Vitiello et al., 2023). Past studies highlight the benefits of interactive and context-sensitive support in group learning (Kumar et al., 2007; Kumar and Rose, 2010). While static scaffolding like fixed prompts (Vogel et al., 2021) and scripted roles (Fischer et al., 2013) have been effective, contextualized interventions within specific conversational contexts (Ai et al., 2010; Cui et al., 2009) or support for student role taking (Gweon; et al., 2007) have also shown positive outcomes. Past studies incorporating dynamic support agents in collaborative learning activities (Kumar et al., 2007; Kumar and Rosé, 2010; Rosé and Ferschke, 2016) have shown the effectiveness of discussion-based learning integrated with conversational support using dialog agents. Finally Sankaranarayanan and colleagues (Sankaranarayanan et al., 2022a; Sankaranarayanan et al., 2022b) have shown the effectiveness of reflection-based learning for collaborative software development, showing that shifting students' focus more towards reflection than actual coding can increase conceptual learning without harming the ability to write code. The contribution of this design paper is the introduction of capabilities from Large Language Models (LLMs) (Vaswani, 2017) to enable new forms of collaborative support agents. While recent studies demonstrate that this new generation of support agents can be effective learning support, the new contribution of this paper is an extension to a publicly available and open-source platform to easily integrate LLM agents developed in the broader CSCL community in order to facilitate needed research to answer questions about how best to use new AI capabilities to support collaborative learning effectively. We provide code for the LLM-bazaar extension, the illustrative instructional example described below, and instructions for obtaining support for using this resource, available on GitHub (Bazaar, 2025).

## GenAI and LLMs for collaborative learning

In this work, we introduce LLM Bazaar, an LLM-empowered agent designed for real-time, multi-agent, multi-party interactions in collaborative learning settings. A challenge that this work addresses is managing open-source LLMs in a way that allows users to specify and configure multiple models with unique persona attributes, such as personality traits and conversational styles, to provide a diverse and contextually nuanced interaction experience. Another challenge is coordinating communication between multiple agents and human users within the same conversational space: each agent must remain aware of recent conversational context to generate relevant responses, which requires a robust context management system that enables agents to share and reference conversation history efficiently. The system must support agent activation, turn-taking, and inter-agent communication to maintain a coherent conversational flow. By supporting flexible, multi-party engagement, our extension to the Bazaar platform creates a responsive learning environment suited to diverse instructional goals and group dynamics. Through this work, we aim to enhance students' engagement in collaborative processes and to provide a foundation for exploring how multi-agent, LLM-empowered interactions impact learning outcomes and collaborative behavior across varied educational scenarios.

## Illustrative learning activity: Regular expressions for recipe organization

This section illustrates how the LLM extension to the Bazaar platform can be applied in real-world educational settings to foster collaboration and provide real-time, context-aware support. We started with a Python programming learning activity we have designed for our own teaching of Python at the college level. The activity is focused on a topic that is particularly challenging for novice programmers, namely regular expressions (regex). As the authors have led collaborative activities for learning regex in their own teaching, the observation has been that this is a topic that lends itself to multiple, creative solution paths and as such affords many opportunities for brainstorming, help-seeking, and knowledge integration.

The activity is set in the JupyterLab infrastructure, which has already been integrated with Bazaar (Sankaranarayanan et al., 2022a,b; Naik et al., 2024). The Jupyter environment allows students to run code cells conveniently to see their programs run and produce output. Additionally, a built-in auto-grader embedded within the notebook is included to check student attempts and offer immediate feedback. In this way, they can test and refine their programs iteratively. This activity has recently been pilot tested in a community college Python course. The collaborative version illustrated here includes improvements based on observations from that pilot and will be deployed in the next iteration of the course.

The learning activity introduces students to the fundamentals of regex through a practical exercise focused on organizing and managing structured text data, specifically using a collection of recipe instructions. Framed as a real-world problem, the activity invites students to help Daphne, a busy home cook, prepare for a family cooking session by using regex to locate, restructure, and optimize her recipe text for quick reference. By working with structured data, students gain experience in building regex patterns to effectively extract, modify, and organize information, which are critical skills in data management and text processing contexts. Table 1 outlines the activity's learning objectives, expected outcomes of the students, and example tasks that we designed to tailor to the corresponding objectives.

**Table 1**
*Learning Objectives of the Regular Expression Activity*

| Learning objectives | Expected outcomes | Example tasks |
|---|---|---|
| Familiarize with regex fundamentals | Understand basic regex syntax and components | overview of regex in a cheat sheet |
| Create regex to match target patterns | Construct and refine regex patterns to identify specific elements in structured text | create a regex pattern for decimal / integer hours (ex. "2 hours") |
| Apply regex to process data | Use regex patterns to extract, modify, and format data for practical use | split the recipe into multiple sections according to the section headers |

To start, we provide students with a foundational overview of regex concepts in the notebook, including key syntax and commonly used functions for locating, splitting, and modifying text. This introduction equips students with the necessary knowledge to approach the series of tasks that we designed to scaffold the learning objectives. The activity itself is divided into eight tasks, each progressively building students' understanding and practical skills with regex. The tasks are organized to reinforce learning through a pattern of creation and application: in one task, students create regex patterns to identify specific information in the recipe text, such as cooking time and section markers; and in the subsequent task, they apply these patterns to extract or modify recipe contents accordingly. For instance, early tasks involve constructing regex patterns to locate integer and decimal hour values within cooking instructions or identifying where each recipe section begins. After identifying these patterns, students modify the text by replacing or splitting it to enhance readability.

Throughout this activity, in a chat window available within the Jupyter environment, the Bazaar architecture allows an LLM-powered agent to connect to the shared workspace in Jupyter and participate in the ongoing discussion and work. The agent listens to the ongoing student conversations and fosters collaboration and discussions by prompting students to share ideas and explore strategies together. If the group still appears to be stuck, the agent provides hints after a preset timer to guide them forward. The Bazaar architecture has been designed to manage the complexity of supporting collaborative groups by distributing the task of listening for particular events in the collaboration and injecting support of various kinds across multiple specialized agents operating collaboratively in the background (Adamson et al., 2014). For example, one agent manages the conversation flow, another maintains student engagement, and a third provides targeted hints. Our extension is to extend the available types of agent support capabilities using LLMs.

## Technical description

LLM Bazaar's key technological advancement is its integration of any number of customizable LLM agents with context-aware interaction support (See architecture in Figure 1). Bazaar itself provides support for collaborative and discussion-based learning, both by directing the learning activity and by facilitating student collaboration through listening and responding to student chat contributions (Adamson et al., 2014). To this, LLM Bazaar adds LLM agents that listen for chat topics related to their assigned prompts. When a relevant topic comes up, the agent responds in a way that fits the flow and content of the conversation.. LLM Bazaar filters potential LLM contributions so that their timing and content is appropriate for the state of the activity.

**Figure 1**
*LLM Bazaar's Integrated Architecture and Information Flow*

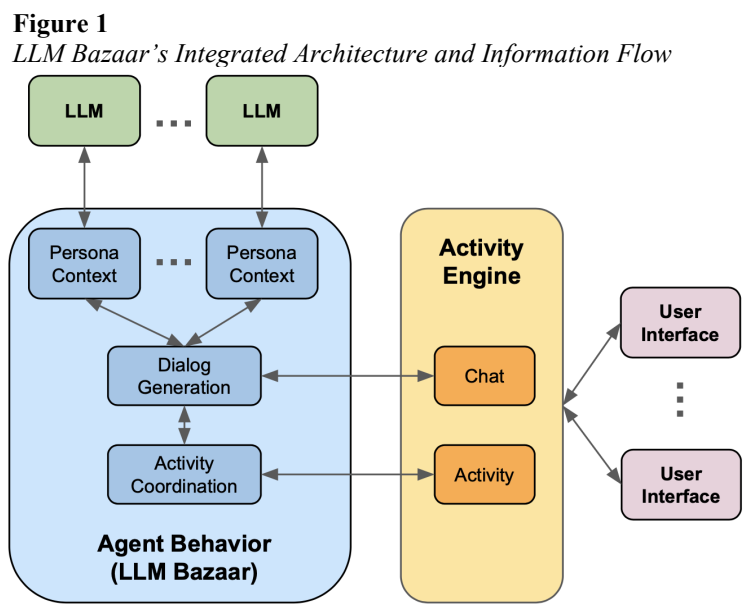

Even without LLM integration, Bazaar exhibits complex behavior: what students perceive as a single entity directing, listening, and responding to the combined state of the activity and the chat space is the emergent behavior of a *plan* along with multiple *listeners*, *actors*, and an *output coordinator*. Bazaar may optionally include an overall *plan* for the activity such as a sequence of subtasks and their timing, along with associated prompts. For flexible behavior during the activity, any number of *listeners* can be configured, each watching for particular events in the activity and/or chat in which they are interested. These range from simple events like noticing a new student presence (or absence) or subtask completion, to more complex ones like noticing opportunities to encourage students to engage in Academically Productive Talk (APT; Michaels et al. 2008) based on student statements that invite APT moves such as elaboration, challenge, or explanation. A listener's assessment of events can be rule-based or the result of a machine-learned classifier. Events that are noticed by any of the listeners are placed on a queue, along with the listener's assessment of the event. Any number of Bazaar's *actors* can be configured to monitor this event queue. Actors optionally propose templated responses to any events on the queue in which they are interested, attaching a priority and a timeout value to each proposal. Examples of actor response proposals include greeting a new student, initializing the next subtask, and responding to opportunities for APT by asking a student to say more, challenging a student statement, or asking for an explanation. With multiple actors proposing responses, there may be multiple competing responses to choose from. That is the role of the *output coordinator*. The output coordinator considers response priorities and timeout values to choose a response, if any, that is appropriate for the current micro and macro state of the activity and chat space. The priority of response proposals may be reduced over time to reflect a shifting focus of attention. In this way, Bazaar performs as a dynamic and subtly flexible collaborative agent even without LLM integration.

LLM Bazaar greatly increases the range of events to which Bazaar can respond, and enhances the creativity of those responses. LLM agents are integrated into Bazaar both as *listeners* for any topics in which they are interested, and as *actors* that propose responses. Bazaar's *output coordinator* maintains ultimate control over which responses, if any, are presented in the activity and chat space. LLMs can respond productively to a wider range of student contributions, such as to help requests on a vast array of topics and to off-topic contributions. In doing so, LLMs can generate well-written, non-templated responses that are situated within the context of the current activity and chat space.

Integrating customized LLM agents within Bazaar is made straightforward by including listener and actor "wrappers" where new LLM agents can be embedded. Plain text configuration files describe both *LLM-specific parameters* (LLM, URL, model type, API key, temperature, etc.) and *personalization parameters* including the learning scenario, instructions about how to respond, and the agent's interest area(s), personality, and conversational style. LLM Bazaar maintains a running history of all chat contributions within the current activity session. For each LLM query, LLM Bazar provides the LLM's personalization parameters and the last *n* chat contributions (where *n* is an LLM configuration parameter, referred to as "context length") to the LLM for context. We will elaborate the configuration design in later sections.

## Example illustrating information flow

Figure 1 provided in the previous section illustrates the flow of information through the system. In this section, we illustrate its operation by displaying how student interactions occur within the integrated environment. The overarching context is provided by the Activity Engine, which in this case is JupyterLab with chat interface enabled. For concreteness, the system will be described below in terms of this activity engine, although LLM Bazaar is by no means constrained to it. See Figure 2 for a screenshot.

**Figure 2**
*Students Working Together with Support from an LLM Agent, Whose Turns are Labeled Bot*

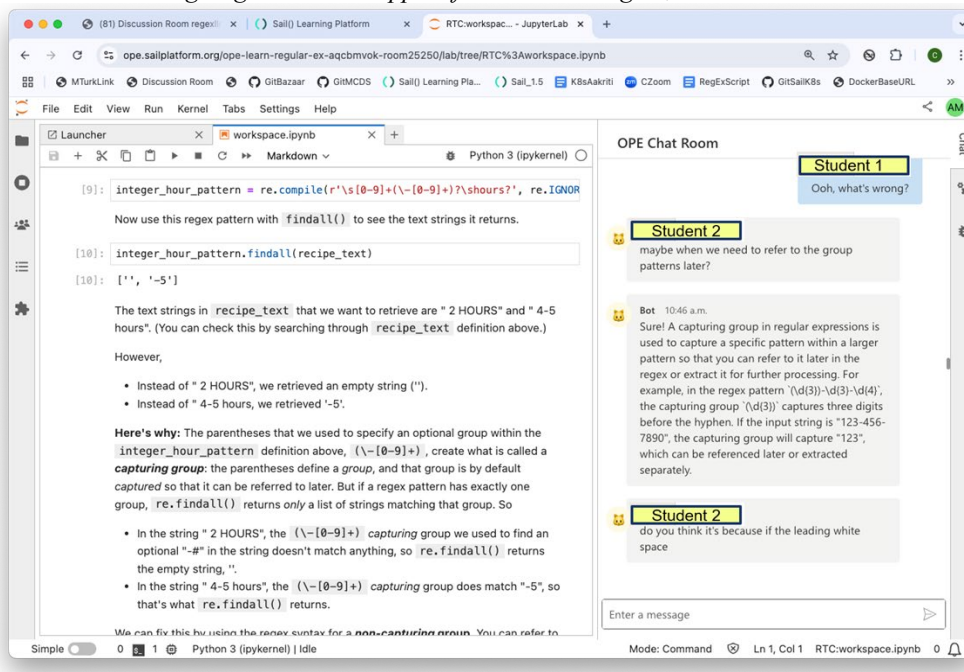

LLM Bazaar uses a *plan* to direct the activity by sending commands to change the activity state – in this case, by adding Jupyter notebook cells to progressively disclose tasks in the students' browser interface – and sending prompts to the chat space. Students work within the Jupyter notebook, reading markdown cells and editing and executing code cells, while communicating with peers through a chat window. Student chat contributions are mostly about how to perform the current task, including help queries, but they also include off-topic contributions such as greetings, exclamations, discussion about other classes, and non-academic pursuits. Within the JupyterLab implementation, these chat contributions are made in the displayed example by typing in a browser window, but Bazaar also enables multimodal interactions – e.g., to support voice contributions (Wang et al., 2020; Vitiello et al., 2023).

Student chat contributions and task state updates (e.g., subtask completion) are forwarded from the activity engine to LLM Bazaar by means of a NodeJS Socket.IO room that is unique to each activity session. Messages received there are examined by multiple listeners, including two LLM listeners, though the number of LLM listeners could easily be reduced or extended. In this activity, the non-LLM listeners look for events like student presence, greetings, and subtask completion. They save all chat contributions both from students and from agents in a chat history store, and optionally queue the event along with the listener's assessment for consideration by LLM Bazaar's actors. Actors, in turn, optionally queue proposed responses for consideration by the output coordinator. The LLM listeners bundle the most recent chat contribution, $n$ turns of chat history, and the LLM personalization parameters, then send the bundle to the associated LLM. Each LLM response is also put on the queue for consideration by the output coordinator. Actions that the output coordinator decides to emit, which may either be chat messages or activity state commands, are sent to the NodeJS room, where they are either displayed in the user interface's chat window or used to update the activity state.

LLM agent behavior is customized using prompt engineering techniques. Prompts designed for two different agent support strategies are displayed in Figure 3 (a) and (b) respectively. Since the regex activity is set in the cooking domain, the prompts cast the agent as a participant in a cooking scenario.

## LLM agent customization and configuration

### Selecting LLMs and setting basic parameters

The initial step in configuring agents involves selecting LLMs and adjusting basic parameters that influence the model's response generation. Our system supports various LLMs, including OpenAI's GPT series and open-source alternatives like Llama models. Once a model is chosen, the system connects to it through a designated API endpoint. This API endpoint serves as a structured interface that facilitates the flow of requests and responses between the system and the model, allowing each agent to generate responses based on user prompts (discussed in the next section) and to respond in real time.

To further refine the model's behavior, users adjust two basic parameters—temperature and context length—within the configuration file. Temperature controls the variability of responses: a higher temperature setting leads to more diverse and creative replies, while a lower one produces more consistent and deterministic outputs (which is a preferable choice for providing tutoring support in our regex activity). Context length, $n$, on the other hand, defines the number of recent conversational turns the model considers when generating a response. By setting the value of $n$ (e.g., 5 turns), users enable agents to reference a relevant and recent segment of the conversation, and track the flow of the discussion and how students are collaborating. With this contextual information, the agent can respond in a way that aligns with the current flow of discussion, selectively providing guidance or encouraging participation as needed. This ensures its contributions are relevant and supportive while maintaining response efficiency by focusing only on the most pertinent portions of the conversation.

### Prompting agents with structured principles

Our configuration file enables users to create custom prompts that ensure agents respond in ways that align with intended instructional goals and desired behaviors. Effective educational interactions require agent responses that are not only clear and relevant but also aligned with specific learning objectives and supportive of group collaboration. This approach is informed by structured prompt design principles adapted from Phoenix and Taylor (2024), which emphasize tailoring prompts to guide agents effectively. By applying these principles, users can ensure agents foster coherent, goal-driven dialogue, encourage teamwork among students, and provide context-aware assistance. We will use the prompt examples from our regex activity in Figure 3 to demonstrate how the principles are applied to prompt our LLM agents effectively.

The first step in designing an effective prompt is to define the behavior, style, and tone each agent should adopt. Users can define a *persona* for each agent, such as name, occupation, age, and personality. These attributes form the agent's "character" and serve as guiding signals for the model's responses. For example, we configure an agent as an "expert programming tutor" who provides step-by-step explanations as they troubleshoot and refine their regex patterns, and encourages active participation from all students. Additionally, users can set a scenario (e.g., a coding lab) to situate conversations in a specific context.

**Figure 3**
*Prompts for Two Agents configured with Distinct Roles to Facilitate the Regex Learning Activity: (a) Providing Tutoring Support for Regex Patterns and (b) Encouraging Participation Among Students*

**Give Direction:** Describe the behavior you want in detail (e.g., style, persona, etc.)

You are an expert programming tutor in a coding lab, helping multiple students learn regular expressions (regex) in a step-by-step, structured manner. Your goal is to guide them through fundamental concepts, provide clear examples, and offer supportive feedback as they progress. Use a friendly, encouraging tone that fosters confidence and clarity.

**Specify Format:** Define rules to follow and the required structure of the response

1. **Start with the basics:** Begin by explaining character classes. Provide simple examples, like matching letters or digits, to illustrate how character classes work. Offer feedback and clarify any misunderstandings before moving on.

2. **Introduce quantifiers:** Explain quantifiers, such as *, +, and {n}. Use straightforward examples to demonstrate how quantifiers modify character classes.

**Divide Labor:** Split complex tasks into steps

3. **Combine concepts for complex patterns:** After covering character classes and quantifiers, guide the students in building more complex regex patterns by combining these elements. Offer step-by-step support as they attempt to create patterns to match more intricate text structures.

4. **Provide targeted feedback:** At each stage, respond with specific feedback based on their progress. Adjust your explanations to their level of understanding and provide hints that help them troubleshoot and refine their regex patterns.

**Provide Examples:** Insert some test cases, like in a coding manual

**Example responses:** If the students struggle with quantifiers, you might respond: "Great effort! Remember, + means 'one or more' of the preceding character. So if you want to match one or more digits, you'd write [0-9]+." Provide similar examples to reinforce learning.

(a)

You are a 45-year-old chef and owner of a trendy cafe who is empathetic, sociable, and loves to teach about cooking. You are in a chat room with several students. You goal is to encourage active participation from all students, ensuring that everyone is contributing to the conversation.

1. **Observe the context:** Pay attention to the ongoing conversation among students. Note the collaboration dynamics, such as who is actively participating and who might need encouragement to contribute.

2. **Encourage participation:** Prompt students to share their ideas. Use open-ended questions and gentle nudges to stimulate discussion.

**Example responses:** If Student A hasn't been actively participating, you might respond: "Let's hear from someone who hasn't spoken yet. Student A, what are your thoughts?"

(b)

In addition, users have options to specify a response structure, such as adjusting response length and level of detail to suit each instructional need. In the regex activity setting, an agent could be set to provide concise hints to inspire thinking and discussion among the group, or elaborate explanations if the students are still stuck after the discussions. To further align responses with instructional goals, users can embed sample responses within the prompt. These examples help guide the LLM to generate responses that meet the desired level of clarity and detail, thus ultimately supporting students more effectively through the learning process.

The principle of dividing labor within the prompt helps the agent handle complex tasks by breaking down instruction into clear steps. For instance, in our regex activity, the prompt could instruct the agent to first cover the basics of character classes with simple examples, then move on to quantifiers, and finally demonstrate how to combine these elements for complex patterns.

Once the initial prompt is set up, evaluating quality becomes an iterative process. Users can use our system to observe the agent's responses in action, refining prompts as needed until the quality of responses aligns with expectations.

### Front end display, feedback, and communication to users

The system's front end is a text-based, web-browser interface that enables users to interact with agents on any standard browser. Each user is assigned a unique ID for session tracking, which helps Bazaar to maintain distinct interactions. The interface is compatible across operating systems, including Windows, macOS, iOS, and Android, providing accessible engagement on various devices. Each session is logged in sequence, enabling analysis to support system refinement and performance monitoring.

## Conclusion and current directions

This paper presents a new resource for the CSCL community, namely LLM Bazaar, which is an extension to the open-source Bazaar architecture, which has been a resource for the community for the past decade (Adamson et al., 2014). Recent advances in AI capabilities available for the community's use call for infrastructure that allows these new capabilities to be easily incorporated into interventions used for this community's ongoing explorations of student interactions with AI and principles for incorporating its capabilities in interventions that support learning. Bespoke demonstrations of these capabilities applied to collaborative learning already exist (Naik et al., 2024). This paper is an invitation to the community to partner in broader investigations of agent designs for LLM-powered support for collaborative learning.

example-based reflection for student learning and transfer task performance. *IEEE Transactions on Learning Technologies*, *15*(5), 594-604.

## Acknowledgments

This work was supported in part by NSF Grants ITEST 2241669, DSES 2222762, and DUE 2100401.